\documentclass[12pt]{article}
\usepackage{amsmath,amssymb,geometry,microtype}
\usepackage{gliederung}
\usepackage{marginnote,xparse,changepage,caption,graphicx,subcaption}

\usepackage{tocloft}

\newcommand{\IGNORE}[1]{}
\newcommand{\be}{\begin{equation}}
\newcommand{\ee}{\end{equation}}
\newcommand{\bea}{\begin{aligned}}
\newcommand{\eea}{\end{aligned}}
\newcommand{\F}[1]{F_{#1}}
\renewcommand{\P}[1]{P_{#1}}
\newcommand{\BF}[1]{\mathbf{F}_{#1}}
\newcommand{\BP}[1]{\mathbf{P}_{#1}}

\newcommand{\ibar}{{\overline{\emph\i}}}
\newcommand{\jbar}{{\overline{\emph\j}}}

\usepackage{todonotes}
\usepackage{soul}

\RequirePackage[colorlinks=true
,urlcolor=blue
,anchorcolor=blue
,citecolor=blue
,filecolor=blue
,linkcolor=blue
,menucolor=blue
,pagecolor=blue
,linktocpage=true
,pdfproducer=medialab
]{hyperref}

\begin{document}

%%%%%%%%%%%%%%%%%%%%%%%%%%%%%%%%%%%%%%%%%% 
%%%%%%%%%%%%%%%%%%%%%%%%%%%%%%%%%%%%%%%%%% 
%%%%%%%%%%%%%%% FOR CQG STYLE %%%%%%%%%%%%%%%%%%
\vfill

\begin{center}
   \baselineskip=16pt {\Huge\bf Boundary Terms for Causal Sets}
  \vskip 1.5cm {Michel Buck${}^a$, Fay Dowker${}^b$,
      Ian Jubb${}^{b}$ and Sumati Surya${}^{c}$}\\
   \vskip .6cm
  \begin{small}
      \textit{
      		${}^{a}${Department of Physics, Northeastern University, Boston, MA 02115, USA}\\\vspace{5pt}
		${}^{b}${Theoretical Physics Group, Blackett Laboratory, Imperial College, London, SW7 2AZ, UK}\\\vspace{3pt}
		${}^{c}${Raman Research Institute, CV Raman Ave, Sadashivanagar, Bangalore 560080, India}
              }
              \end{small}\\*[.6cm]
   \end{center}

\vspace{20pt}

\begin{center}
\textbf{Abstract}
\end{center}

\begin{quote}
We propose a family of boundary terms for the action of a causal set with a spacelike boundary. 
We show that in the continuum limit one recovers the Gibbons-Hawking-York boundary term in the mean. 
We also calculate the continuum limit of the mean causal set action for an Alexandrov interval in flat spacetime.
We find that it is equal to the volume of the codimension-2 intersection of the two light-cone boundaries of the interval. 
\end{quote}

%%%%%%%%%%%%%%%%%% FOR CQG %%%%%%%%%%%%%%%%%% 
%%%%%%%%%%%%%%%%%%%%%%%%%%%%%%%%%%%%%%%%%% 
%%%%%%%%%%%%%%%%%%%%%%%%%%%%%%%%%%%%%%%%%% 
\pagebreak
\tableofcontents

\section{Introduction}

One approach to constructing a quantum dynamics for the causal set approach to quantum gravity \cite{Bombelli:1987aa}
is to discover a discrete counterpart of the gravitational action, $S[\mathcal C]$  that can furnish the weight,  $e^{iS[\mathcal C]}$,   of each causal set, $\mathcal C$, 
in the gravitational sum over histories.  A start in this direction has been made with a
 proposal for scalar curvature estimators for causal sets of dimension $d$ 
  \cite{Benincasa_Dowker:The_Scalar_Curvature_of_a_Causal_Set, Dowker_Glaser:dAlembertians_for_Causal_Sets, Glaser:2013xha}.
 Summing such a scalar curvature estimator over all elements of a causal set (\textit{causet} for short) gives a natural 
 proposal for a causet analogue of the Einstein-Hilbert action, a proposal that remains to be 
 studied in depth. The purpose of this paper is to 
 initiate an investigation of gravitational boundary terms for the action of causets. This is likely to be important as, 
 in the continuum, it is well known that the Einstein-Hilbert action, $S_{EH}$, is not the full story in the presence of spacetime boundaries. 
 Indeed, the gravitational action must include a boundary term $S_{GHY}$, the Gibbons-Hawking-York (GHY) boundary term, 
in order to yield a well-defined variational principle when the metric is fixed 
on the boundary of spacetime~\cite{York:1972, Gibbons_Hawking_Boundary}. If the classical limit of quantum gravity
is to arise from the path integral in the expected way, 
such a term in the action will be essential when boundaries are 
present. Whilst we do not yet know how to fix boundary conditions for 
causets in general, it is  likely to be useful to have an analogue 
of the GHY boundary term for any causal set which is well-approximated by a manifold with a boundary. In this 
paper we initiate the study of boundary terms for causal sets by proposing an analogue in the case of spacelike boundaries and investigating the above mentioned causal set action for a causal interval with null boundaries. 
\sloppy
First we consider causal sets which are well approximated by $(M,g)$, a
$d$-dimensional, causal, Lorentzian  spacetime with finite volume which
admits a closed, compact spacelike submanifold, $\Sigma$, such that the causal past and future sets, $M^\pm:=J^\pm (\Sigma)$, satisfy $M^+\cap M^-=\Sigma$. 
Then $\Sigma$ is a component of the future (past)  spacelike boundary for $M^-$ ($M^+$) and
the GHY term for $\Sigma$, considered as a boundary of $M^+$ or $M^-$, is given by
\be\label{eq:GHYBT_in_continuum}
{S}_{GHY}\left[\Sigma, M^\pm\right]= \mp \frac{1}{l_p^{d-2}}\int_{\Sigma} d^{d-1}x\: \sqrt{h}\, K \:,
\ee
where $K$ is the trace of the extrinsic curvature $K_{\mu\nu}=h_{\mu}^\rho h_\nu^\sigma \nabla_\rho n_\sigma$ of $\Sigma$, $l_p = (8\pi G)^{\frac{1}{d-2}}$ is the rationalised Planck length and we are working in units 
where $\hbar=1$. Here we take $n_{\mu}$ to be the future-pointing timelike unit covector normal to $\Sigma$. We work with a mostly plus convention for the metric so 
 $n^{\mu}$ is past-pointing. 

We recall that the integral in~\eqref{eq:GHYBT_in_continuum} is equal to the normal derivative of the volume of $\Sigma$ along the unit normal vector field, $n^{\mu}$:
\be\label{eq:normal_deriv_boundary}
\int_\Sigma d^{d-1}x\: \sqrt{h}\, K = \frac{\partial}{\partial n}\int_\Sigma d^{d-1}x\: \sqrt{h} \;,
\ee
where this is the rate of change of the volume backwards in time, as $n^{\mu}$ is past-pointing. This observation suggests a natural candidate for an analogue of the GHY boundary term for $\Sigma$ for a causet that can be faithfully embedded in $M$.  Spacetime volume corresponds
to cardinality in a causet. Hence the spatial volume gradient corresponds intuitively to the difference between the number of causet elements that are future nearest neighbours of $\Sigma$ and the number of past nearest neighbours. 
This intuition turns out to be a good guide and we will identify a family of causal set boundary terms based on it. The family of causet boundary terms we find corresponds to the 
different ways to define a discrete derivative that tend to the same limit in the 
continuum. We will define the causet functions and state the claims
about their continuum limits in section \ref{claims}. The calculations 
justifying the claims appear in section \ref{calcs}. In section \ref{intervals} we investigate the proposed causal set 
action for causal intervals in flat spacetime and show that its mean takes the form, in the 
continuum limit, of a boundary contribution from the codimension 2 ``joint'' of the 
interval's boundary.

\section{The Boundary Terms}\label{claims}

Given a finite causet $(\mathcal D, \preceq)$, we define an
``$\F{k}$ element'' to be an element of $\mathcal{D}$ with exactly $k$ elements -- not 
equal to itself -- to its future in the order\footnote{Other ways to 
express this are, ``has exactly $k$ descendants'' and ``precedes exactly $k$ elements 
excluding itself''.} and define  $\F{k}\left[\mathcal{D}\,\right]$ to be the number of 
$\F{k}$ elements in $\mathcal D$. 
Similarly a ``$\P{k}$ element'' is an element  with exactly $k$ elements to its past in the 
order and  
  $\P{k}\left[\mathcal{D}\,\right]$ is the number of $\P{k}$ elements 
 in $\mathcal{D}$. 
For example, $F_0[\mathcal{D}]$ ($P_0[\mathcal{D}]$) is the number of 
maximal (minimal) elements of $\mathcal{D}$. 

Given a finite causet, $(\mathcal{C},\preceq)$, with  two subcausets, ${\mathcal{C}}^+$ and ${\mathcal{C}}^-$ 
we introduce the following family of  causal set ``boundary terms'' (CBT):
\be\label{general_boundary_sum}
S^{ (d)}_{CBT}\left[\mathcal{C}, \mathcal{C}^-,\mathcal{C}^+;\vec{p}, \vec{q}\,\right]:= \left (l/l_p\right)^{d-2} a_{d}
\left ( \sum_m p_m \F{m}\left[\mathcal{C}^- \right]
+  \sum_n q_n \P{n}\left[\mathcal{C}^+ \right]\right) \;,
\ee
where the constant $a_{d}$ is given by
\be\label{Cn}
a_{d}=\frac{d (d+1)}{ (d+2)}\left(\frac{S_{d-2}}{d(d-1)}\right)^{\frac{2}{d}} \;.
\ee
$S_d=(d+1)\pi^{\frac{d+1}{2}}/\Gamma\left (\frac{d+1}{2}+1\right)$ is the volume of the unit $d$-sphere,
 $l$ is a length and $l_p$ the Planck length  defined previously.
$\vec{p}$ and $\vec{q}$ denote finite strings of real numbers $ (p_0,\ldots,p_m,\ldots)$ and $ (q_0,\ldots,q_n,\ldots)$ respectively. We will prove that the strings must satisfy the following conditions:
\begin{align}\label{coefficient_relation1}
& \sum_m p_m \frac{\Gamma\left (\frac{1}{d}+m \right)}{m!}  + \sum_n q_n\frac{\Gamma\left (\frac{1}{d}+n \right)}{n!}=0 \;,
\\
& \label{coefficient_relation2}\sum_m p_m \frac{\Gamma\left (\frac{2}{d}+m \right)}{m!}  - \sum_n q_n\frac{\Gamma\left (\frac{2}{d}+n \right)}{n!}=1 \;.
\end{align}

We call \eqref{general_boundary_sum} a boundary term but in general, when 
${\mathcal{C}}^+$ and ${\mathcal{C}}^-$ are arbitrary subcausets of $\mathcal{C}$, it will have no
physical significance.

Let $(M,g)$ be a $d$-dimensional spacetime with
 finite volume and spacelike, closed, compact hypersurface $\Sigma$ as 
 described in the introduction. 
Given such a spacetime $(M,g)$ and sets $M^\pm := J^\pm(\Sigma)$, 
 $S^{ (d)}_{CBT}$  defines a family of random variables in the following way. The Poisson 
 process of sprinkling points into $M$ with density $\rho=l^{-d}$ generates a random causet $(\mathcal{C},\preceq)$ together with subcausets $\mathcal{C}^\pm$ which consist of those elements sprinkled into $M^\pm$. The functions $\P{k}$
  and $\F{k}$ acting on the random causets $\mathcal{C}^+$ and  $\mathcal{C}^-$ 
 respectively are random variables $\BP{k}$ and $\BF{k}$. 
These random variables can be substituted into~\eqref{general_boundary_sum} 
to give the family of random variables $\textbf{S}^{ (d)}_{CBT}$: 
 \be\label{random_bt}
\textbf{S}^{ (d)}_{CBT}\left[M, \Sigma,\rho; \vec{p}, \vec{q}\,\right]:= \left (l/l_p\right)^{d-2} a_{d}
\left ( \sum_m p_m \BF{m}+  \sum_n q_n \BP{n}\right) \;.
\ee
 We claim that in the limit of infinite density the expectation value, in the sprinkling process, of $\textbf{S}^{ (d)}_{CBT}$ tends to the continuum GHY boundary term of the surface $\Sigma$:
\be
\lim_{l\rightarrow0}\left\langle\textbf{S}^{ (d)}_{CBT}[M, \Sigma,\rho;\vec{p} , \vec{q}]\right\rangle= \frac{1}{l_p^{d-2}}\int_{\Sigma} d^{d-1}x\: \sqrt{h}\: K = {S}_{GHY}\left[\Sigma, M^-\right] \;,\label{eq:mainconjecture}
\ee
where $\langle\cdot\rangle$ denotes the mean over sprinklings. We will prove this in the next section.

 One can see that at least two non-zero entries in $\vec{p}$ and $\vec{q}$
 together are necessary in order to satisfy~\eqref{coefficient_relation1} and~\eqref{coefficient_relation2} and 
 if exactly two entries are non-zero they will be uniquely fixed, but if more than two entries are non-zero this uniqueness is lost.  This accords with the continuum boundary term  being 
 a first derivative. The freedom of choice in $\vec{p}$ and $\vec{q}$ is the freedom to discretise a derivative in many ways but the difference of two nearby values is sufficient. 

\begin{figure}
  \centering
    {\includegraphics[width=\textwidth]{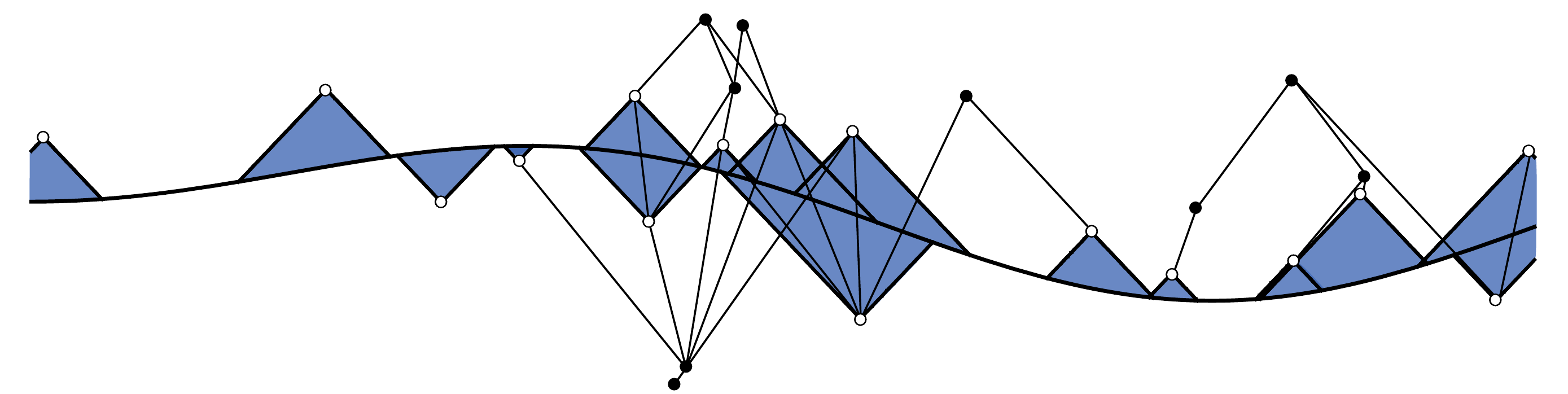}}
     \caption{An illustration of a sprinkling into a spacetime partitioned by a spacelike hypersurface. Black points correspond to causal set elements and links (irreducible causal relations) between elements are shown as thin black lines. The maximal ($F_0$) elements in $\mathcal C^-$ and the minimal ($P_0$) elements in $\mathcal C^+$ have been highlighted with white filling. The shaded areas illustrate the regions whose volumes $V_\blacktriangle$ and $V_\blacktriangledown$ are needed in the proof of Section~\ref{calcs}. In this sketch $P_0=11$ and $F_0=5$ (with time flowing upwards).}
     \label{fig:Nmin_Nmax}
\end{figure}

We introduce special notation for the simplest member of the family: 
\be\label{eq:min_max_simple_formula}
S^{ (d)}_{0}\left[\mathcal{C}, \mathcal{C}^-,\mathcal{C}^+\right]:=
\left (l/l_p\right)^{d-2}\frac{a_{d}}{2\Gamma\left (\frac{2}{d} \right)}\left ( \F{0}\left[\mathcal{C}^- \right] - \P{0}\left[\mathcal{C}^+ \right] \right) \;.
\ee
This is proportional to the difference in the numbers of minimal elements of $\mathcal{C}^+$ and maximal elements of $\mathcal{C}^-$, and thus corresponds to the intuitive idea described in the introduction. An illustrative sketch of the idea is shown in Figure~\ref{fig:Nmin_Nmax}. This case is the easiest to investigate computationally, and we shall use its random variable counterpart, $\textbf{S}^{ (d)}_{0}\left[M,\Sigma,\rho\right]$, later when we study the fluctuations of the discrete boundary terms numerically.

There are two special subfamilies of boundary terms, one defined by $\vec{p} = 0$ and the other by 
$\vec{q} = 0$. In the former (latter) case, this corresponds to defining a boundary term for the 
past (future) boundary of $\mathcal{C}^+$  ($\mathcal{C}^-$) using only data from $\mathcal{C}^+$ 
($\mathcal{C}^-$) itself. The simplest cases of these boundary terms are 
\begin{align}
\label{eq:future_past_boundary_terms2}
S^{ (d)}_{-}[\mathcal{C}^+]:= & \left (l/l_p\right)^{d-2}\frac{a_{d}}{\Gamma\left (\frac{2}{d} \right)}\left ( \P{0}\left[\mathcal{C}^+\,\right] - d\: \P{1}\left[\mathcal{C}^+\,\right] \right) \;,\\
\label{eq:future_past_boundary_terms1}
S^{ (d)}_{+}[\mathcal{C}^-]:= &\left (l/l_p\right)^{d-2}\frac{a_{d}}{\Gamma\left (\frac{2}{d} \right)}\left ( d\:\F{1}\left[\mathcal{C}^-\,\right] - \F{0}\left[\mathcal{C}^-\,\right] \right)  \;.
\end{align}
These give rise to random variables $\textbf{S}^{ (d)}_{-}[M, \Sigma, \rho]$ 
and $\textbf{S}^{ (d)}_{+}[M, \Sigma, \rho]$ via sprinkling at density $\rho = l^{-d}$ as before.

\subsection{The Surface Volume Family}

We also propose a family of causet functions that will give the volume of a spacelike 
hypersurface in the appropriate context:
\be\label{general_area_sum}
A^{ (d)}[\mathcal{C}, \mathcal{C}^-,\mathcal{C}^+;\vec{p},\vec{q}]:=\left (l/l_p\right)^{d-1}b_{d}\left (\sum_m p_m \F{m}\left[\mathcal{C}^- \right] + \sum_n q_n \P{n}\left[\mathcal{C}^+ \right]\right) \;,
\ee
where 
\be\label{constant_b_d}
b_d=d\left(\frac{S_{d-2}}{d(d-1)}\right)^{\frac{1}{d}} \;,
\ee
and  $\vec{p}$ and $\vec{q}$ now satisfy 
\be\label{area_coefficient_relation}
\sum_m p_m \frac{\Gamma\left (\frac{1}{d}+m \right)}{m!}  + \sum_n q_n\frac{\Gamma\left (\frac{1}{d}+n \right)}{n!}=1 \;.
\ee
We see that only one non-zero entry is necessary to give an expression for the discrete surface volume. Once again, for $(M,g)$, $\Sigma$ and $\rho = l^{-d}$,
we can define a family of random variables,
 \be\label{randomvol}
\textbf{A}^{ (d)}\left[M, \Sigma,\rho; \vec{p}, \vec{q}\,\right]:= \left (l/l_p\right)^{d-2} b_{d}
\left ( \sum_m p_m \BF{m}+  \sum_n q_n \BP{n}\right) \;.
\ee
We claim that, in the limit of infinite density, the expectation value of $\textbf{A}^{ (d)}$ in the sprinkling process tends to the spatial volume of the surface $\Sigma$:
\be\label{eq:conjecture_for_area}
\lim_{l\rightarrow0}\left\langle\textbf{A}^{ (d)}[M,\Sigma, \rho;\vec{p} , \vec{q}]\right\rangle= \frac{1}{l_p^{d-1}}\int_{\Sigma} d^{d-1}x\: \sqrt{h} \;.
\ee
In proving~\eqref{eq:mainconjecture} we will establish the results necessary to prove this claim as well.

One can define functions for the volumes of future and past boundaries respectively as the two simplest members of the family:
\begin{align}\label{eq:future_past_spatial_volume}
A^{ (d)}_{+}[\mathcal{C}^-]:=& \left (l/l_p\right)^{d-1}\frac{b_{d}}{\Gamma\left (\frac{1}{d}\right)}\: \F{0}[\mathcal{C}^-] \;,
\\
A^{ (d)}_{-}[\mathcal{C}^+]:= & \left (l/l_p\right)^{d-1}\frac{b_{d}}{\Gamma\left (\frac{1}{d}\right)}\: \P{0}[\mathcal{C}^+] \;.
\end{align}

\section{Calculations}\label{calcs}

\subsection{$\left\langle \BF{m}\right\rangle$ and $\left\langle \BP{m}\right\rangle$ for Poisson Sprinklings}

In order to establish~\eqref{eq:mainconjecture} and~\eqref{eq:conjecture_for_area} we will find the behaviour of the mean values of $\BF{m}$ and $\BP{m}$ as $\rho$ tends to 
infinity ($l \rightarrow 0$). The result will follow almost immediately.

For any given realisation of the sprinkling, the probability that a sprinkled point $p\in M^-$ below the surface $\Sigma$ is an $F_k$ element of $M^-$ is given by the probability that $k$ points of the sprinkling lie in the region $J^{+} (p)\cap J^{-} (\Sigma)$. The Poisson process assigns a probability,
\be\label{Poisson}
\mathbb P\left (\text{k points in }J^{+} (p)\cap J^{-} (\Sigma)\right)=\frac{\left (\rho\: V_\blacktriangledown (p)\right)^k}{k!}e^{-\rho V_\blacktriangledown (p)} \;,
\ee
to this event, where $V_\blacktriangledown (p):= \mathrm{vol} (J^{+} (p)\cap J^{-} (\Sigma))$ is the spacetime volume of the region $J^{+} (p)\cap J^{-} (\Sigma)$. The probability of sprinkling an element into an infinitesimal $d$-volume $dV_p$ at $p$ is $\rho dV_p$ where $\rho=l^{-d}$ is the sprinkling density, and so the total expected number of $\F{k}$ elements below $\Sigma$ is
\be\label{eq:nmax}
\left\langle \BF{k}\right\rangle =\rho\int_{J^{-} (\Sigma)}dV_p\; \frac{\left (\rho\: V_\blacktriangledown (p)\right)^k}{k!}e^{-\rho V_\blacktriangledown (p)} \;.
\ee
Similarly the expected number of $\P{k}$ elements above $\Sigma$ is
\be\label{eq:nmin}
\left\langle \BP{k}\right\rangle =\rho\int_{J^{+} (\Sigma)}dV_p\; \frac{\left (\rho\: V_\blacktriangle (p)\right)^k}{k!}e^{-\rho V_\blacktriangle (p)} \;,
\ee
where $V_\blacktriangle (p):= \textrm{vol} (J^{+} (\Sigma)\cap J^{-} (p))$.

Let  $x^\mu= (t,\mathbf x)$ be ``synchronous'' or Gaussian Normal Coordinates (GNCs) adapted to $\Sigma$ such that in a neighbourhood $U_\Sigma$ of $\Sigma$ the line element is
\be
ds^2 = -dt^2 + h_{ij} (t,\mathbf x) dx^i dx^j \;.
\ee
In these coordinates the surface $\Sigma$ corresponds to $t=0$. Each point $p$ in the neighbourhood lies on a unique timelike geodesic whose tangent vector at $\Sigma$ is equal to $-n^\mu$ on $\Sigma$, and the $t$ coordinate of $p$  is equal to the proper time from $\Sigma$ to $p$ along that geodesic.
The spacetime $({M}, g)$ restricted to this neighbourhood of $\Sigma$ is globally 
hyperbolic and $\Sigma$ is a Cauchy surface within it. 

The integrals~\eqref{eq:nmax} and~\eqref{eq:nmin} seem intractable as they stand, since the integration is over the entire causal past/future of the surface. However, since $\Sigma$ is closed and compact and $M^+$ and $M^-$ are of finite volume, we can always find a subneighbourhood of $U_\Sigma$ such that the contribution to the integrals from the complement of that subneighbourhood tends to zero exponentially quickly as $\rho\rightarrow\infty$. Let $\varepsilon > 0$ be small enough such that for all $p\in\Sigma$ and $|t|<\varepsilon$, $(t,\mathbf{x}(p))$ are the GNCs of a point in $U_{\Sigma}$. Define $U_{\Sigma}(\varepsilon):=\lbrace q\in U_{\Sigma}:|t(q)|<\varepsilon \rbrace$ and consider the integral in \eqref{eq:nmin}  restricted to 
$W:= J^+(\Sigma) \setminus U_\Sigma(\varepsilon)$:
\be\label{eq:expfalloff}
\Big| \, \int_{W}dV_p\; \frac{\left (\rho\: V_\blacktriangle (p)\right)^k}{k!}e^{-\rho V_\blacktriangle (p)}\Big| \le \| e^{-\rho V_\blacktriangle }\| \int_{W}dV_p\; \frac{\left (\rho\: V_\blacktriangle (p)\right)^k}{k!} \;,
\ee
where $\| e^{-\rho V_\blacktriangle }\|$ is the uniform norm over the integration region $W$. 
Since $V_\blacktriangle (p)$ increases with $t$ along the geodesics from $\Sigma$, and 
$\{ x^\mu \in U_\Sigma \,:\, t= \varepsilon\}$ is homeomorphic to $\Sigma$ and so is closed and 
compact,  
$ V_\blacktriangle (p)$ achieves its minimum value $V_{min}> 0$ in $W$ at some point 
with $t =\varepsilon$. Then $\| e^{-\rho V_\blacktriangle}\| = e^{-\rho V_{min}}$ and 
so the integral \eqref{eq:expfalloff} falls off exponentially fast as $\rho \rightarrow \infty$. 
Similarly for \eqref{eq:nmax}.

Thus, so long as $\rho$ is large enough, we make only an exponentially small 
error by cutting off
the integration ranges in \eqref{eq:nmax} and \eqref{eq:nmin} at 
$t=\pm\varepsilon$ with $\varepsilon$ as small as we need 
in order to be able to expand in powers of $t$. Expanding the determinant of the
metric around $t=0$, the integrals we want to evaluate are 
\begin{gather}\label{eq:nmax_and_eq:nmin}
\begin{aligned}
\left\langle \BF{k}\right\rangle & = \rho \int_{\Sigma}d^{d-1}x\int_{-\varepsilon}^{0}dt\:
h^{\frac{1}{2}}\left (1+
\frac{1}{2}\frac{\dot{h}}{h}t+O (t^2)\right)
 \frac{\left (\rho\: V_\blacktriangledown (p)\right)^k}{k!} e^{-\rho V_\blacktriangledown (t,\mathbf x)} +\dots \;,
\\
\left\langle \BP{k}\right\rangle & = \rho \int_{\Sigma}d^{d-1}x\int_{0}^{\varepsilon}dt\:
h^{\frac{1}{2}}\left (1+
\frac{1}{2}\frac{\dot{h}}{h}t+O (t^2)\right) \frac{\left (\rho\: V_\blacktriangle (p)\right)^k}{k!} e^{-\rho V_\blacktriangle (t,\mathbf x)} +\dots \;,
\end{aligned}
\end{gather}
where both $h:= det\left (h_{ij}\right)$ and $\dot{h}\: := \frac{\partial h}{\partial t}$ 
are evaluated at $t=0$, 
and $+ \dots$ denotes  ``terms that vanish exponentially fast in the limit $\rho \rightarrow \infty$''.

\subsection{Volumes}

Let us first calculate $V_\blacktriangle (p)$, the volume of region $\mathcal{X}_p: = J^-(p) \cap M^+$, for $p$ in $U_\Sigma(\varepsilon)\cap M^+$. Let $p_0$ be the point on $\Sigma$ where the unique timelike geodesic through $p$ whose tangent is normal to $\Sigma$ intersects $\Sigma$. 
Let the values of $p$'s GNCs be
$x^\mu_p= (t_p,\mathbf x_p)$, then $p_0$ has GNCs $x_0^\mu:=x^\mu (p_0)= (0,\mathbf x_p)$.
We choose $\varepsilon$ small enough that, for every $p\in U_\Sigma(\varepsilon)\cap M^+$, there exists a Riemann normal neighbourhood centred on $p_0$ containing the region $J^-(p) \cap M^+$. We choose Riemann Normal Coordinates (RNCs) centered at $p_0$, $y^{\overline{\mu}}=(y^{\overline{0}},\mathbf{y})=(\overline{t},\mathbf{y})$, such that the GNC time coordinate of $p$ equals the RNC time coordinate of $p$:  $t_p = \overline{t}_p =: T$. 

The relationship between RNCs  $y^{\overline\mu}$ and GNCs $x^\nu$ is, to second order, 
\be\label{eq:RNCtotaltrans}
y^{\overline{\mu}}=A^{\overline{\mu}}_{\;\nu}(x^\nu -x_0^{\nu})+\frac{1}{2}A^{\overline{\mu}}_{\;\mu}\Gamma^{\mu}_{\;\nu\rho} (p_0)(x^\nu -x_0^{\nu}) (x^\rho -x_0^{\rho})+O ( (x-x_0)^3) \;.
\ee
The constant matrix $A^{\overline\mu}_{\;\mu}$ obeys 
\be\label{eq:RNCMetricTransAtPAndChris}
A^{\overline\mu}_{\;\mu} A^{\overline\nu}_{\;\nu}\eta_{\overline\mu\overline\nu}=g_{\mu\nu}(p_0) \;,
\ee
and the metric and Christoffel symbols in RNCs are flat at $p_0$:
\be\bea
g_{\overline{\mu} \overline{\nu}} (p_0)&=\eta_{\overline\mu\overline\nu} \;,\\\Gamma^{\overline{\mu}}_{\;\overline{\nu}\overline{\rho}} (p_0)&=0 \;.
\eea\ee
The inverse coordinate transformation is
\be\label{eq:RNCinversetrans}
x^{\mu}=x_0^{\mu}+A^{\mu}_{\;\overline{\mu}}y^{\overline{\mu}}+O (y^3) \;,
\ee
where $A^{\mu}_{\;\overline{\mu}}$ is the inverse matrix of $A^{\overline\mu}_{\;{\mu}}$, i.e. $A^{\overline{\mu}}_{\;\mu}A^{\mu}_{\;\overline{\nu}}=\delta^{\overline{\mu}}_{\;\overline{\nu}}$ and $A^{\mu}_{\;\overline{\mu}}A^{\overline{\mu}}_{\;\nu}=\delta^{\mu}_{\;\nu}$. There is no $O(y^2)$ term in~\eqref{eq:RNCinversetrans} due to the fact that $\Gamma^{\overline{\mu}}_{\;\overline{\nu}\overline{\rho}} (p_0)=0$. The
 components of $A^{\overline{\mu}}_{\mu}$ satisfy $A^{\overline 0}_{\;0}=1$, $A^{\overline 0}_{\;i}=0$, $A^{\ibar}_{\;i}A^{\jbar}_{\;j}\delta_{\ibar\jbar} =h_{ij}(p_0)$ and $\delta_{\ibar\jbar}=A^i_{\;\ibar}A^j_{\;\jbar}h_{ij} (p_0)$.
 
As we saw above, the limit we are considering, $\rho \rightarrow \infty$,
allows us to take $\varepsilon$ to be arbitrarily small so the limit
can be thought of as driving each relevant $p \rightarrow p_0$ on $\Sigma$ along the geodesic
normal to $\Sigma$. That makes the region $\mathcal{X}_p$,
whose volume we need, tend to a truncated solid, nearly flat cone with apex $p$ and 
a base on $\Sigma$ defined by a 
quadratic form in the three spatial RNCs around $p_0$. The leading contribution to the volume, 
\be\label{eq:VolumeWithNoSimplifications}
V_\blacktriangle (p)=\int_{\mathcal{X}_p} d^d y\;\sqrt{-g(y)} \;,
\ee
is therefore the volume, $\frac{S_{d-2}}{d(d-1)}T^d$, 
of the flat cone of height $T$ with a flat base on surface $\overline{t} = 0$. Corrections to this are higher order in $T$ and come from three sources: \textit{(i)}  $\sqrt{-g(y)} \ne 1$, \textit{(ii)} the null geodesics down from $p$ to $\Sigma$ forming the top boundary, $T_p:=\partial J^-(p)\cap M^+$, of $\mathcal{X}_p$ 
are not straight, and \textit{(iii)} the base, $B_p: =\Sigma\cap J^-(p)$ is not a flat disc. 
The first two corrections are due to the curvature of $M$ and the third comes from the extrinsic 
curvature of $\Sigma$. 

The correction from \textit{(iii)} is found by taking spacetime to be flat, so that RNCs are 
the usual Cartesian coordinates centred at $p_0$ and  $T_p$ is the top boundary of the
flat cone, satisfying $\sum_{\emph\i=1}^{d-1}\left(y^\ibar\right)^2= (T-\overline{t})^2$ and $\overline t \in [0,T]$.
The base $B_p$ in GNCs lies in surface $t=0$, so we can use (\ref{eq:RNCtotaltrans}) to find the equation for the surface in RNCs. This gives
\be\label{eq:BottomSurfaceWithGNC}
\overline{t}=\frac{1}{2}\Gamma^{0}_{\;ij} (p_0)(x^i-x_0^i) (x^j-x_0^j)+O ( (x-x_0)^3) \;.
\ee
The linear part on the right hand side of (\ref{eq:RNCtotaltrans}) vanishes, since  $A^{\overline{0}}_{\;\mu}(x^{\mu}-x_0^{\mu})=t$ (which follows from  $A^{\overline{0}}_{\;i}=0$ and $A^{\overline{0}}_{\;0}=1$) and $t=0$ on the bottom surface. Using the inverse RNC relation (\ref{eq:RNCinversetrans}), the equation for $B_p$ in RNCs is 
\be\label{eq:BottomSurface}
\overline{t}=\frac{1}{2}\Gamma^{0}_{\;ij} (p_0)A^{i}_{\;\ibar}A^{j}_{\;\jbar}y^{\ibar} y^{\jbar}+O (y^3) \;.
\ee
Let us rewrite this equation in spherical polar coordinates, i.e. define $r:=\sqrt{\delta_{\ibar\jbar}y^\ibar y^\jbar}$ and the usual angular coordinates $\phi_1,..,\phi_{d-2}$ in terms of the spatial coordinates $y^{\overline{1}} = r \cos (\phi_1),\ldots, y^{\overline{d-1}} = r \sin (\phi_1) \cdots \sin (\phi_{d-3}) \sin (\phi_{d-2})$. Then
\be\label{eq:RadialBottomSurface}
\overline{t}=\frac{1}{2}\left (\Gamma^{0}_{\;ij} (p_0)A^{i}_{\;\ibar}A^{j}_{\;\jbar}\frac{y^{\ibar} y^{\jbar}}{r^2}\right)r^2+O (y^3)=\frac{1}{2}f (\mathbf{x}_p,{\boldsymbol\phi})r^2+O (y^3) \;,
\ee
where $\boldsymbol\phi$ stands collectively for all the angular coordinates $\phi_1,..,\phi_{d-2}$. The function $f (\mathbf{x}_p,\boldsymbol\phi)$ depends on $\mathbf{x}_p$ since $\Gamma^{0}_{\;ij}$ and $A^{i}_{\;\ibar}$ depend on $p_0$.

\begin{figure}[t]
  \centering
    {\includegraphics[scale=0.5,trim={1cm 5cm 1cm 6cm},clip]{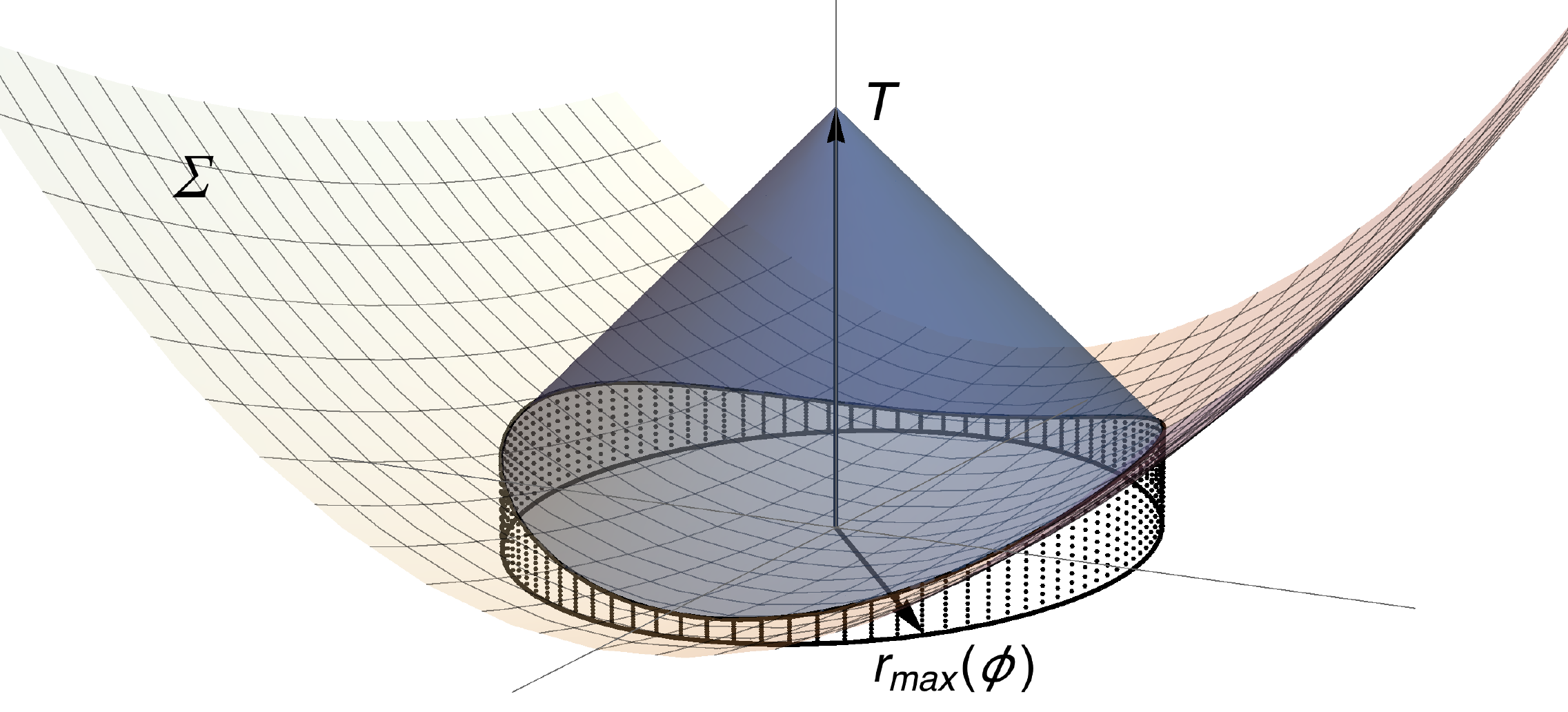}}
     \caption{A $3$-dimensional representation of the region $\mathcal{X}_p$ in RNCs. The top, $T_p$, of $\partial\mathcal{X}_p$ can be approximated as a flat cone, and the base, $B_p$, intersects 
   $T_p$ at a radial coordinate, $r_{max} (\boldsymbol\phi)$, which will in general be a function of the angles $\boldsymbol\phi$ (in $3$ dimensions there is one angle $\phi$). This function is found by projecting down from the intersection to the $\overline{t}$ plane.   }  \label{fig:cone_plot}
\end{figure}

With the boundaries of the integration region in hand, we can now write down the integral explicitly in spherical coordinates:
\be\label{eq:VolumeIntegralSpherical}
\int_{\mathcal{X}_p \ \textrm{in flat space}} d^d y
=\int_{S^{d-2}}
d\Omega_{d-2}
\int_{0}^{r_{max} (\phi)}r^{d-2}dr
\int_{\frac{1}{2}f (\mathbf{x}_p,\phi)r^2}^{-r+T}
d\overline{t}+O (T^{d+2}) \;,
\ee
where $r_{max} (\boldsymbol\phi)$ is the value of the radial coordinate for which $B_p$  intersects $T_p$ at an angle $\boldsymbol\phi$, as shown in Figure \ref{fig:cone_plot}. Equating the time coordinates of $T_p$ and $B_p$ gives 
\be
\frac{1}{2}f (\mathbf{x}_p,\boldsymbol\phi){r_{max}}^2 (\boldsymbol\phi)=-r_{max} (\boldsymbol\phi)+T \;.
\ee
We
solve this for $r_{max} (\boldsymbol\phi)$ and take the positive solution. The solution can be expanded in $T$ and is simply $r_{max}=T+O (T^2)$, with angular dependent terms contributing at $O (T^2)$. The $O (T^2)$ term will contribute at $O (T^{d+2})$ in the volume integral. Substituting $r_{max}=T$ into (\ref{eq:VolumeIntegralSpherical}) allows us to evaluate the integral \eqref{eq:VolumeIntegralSpherical}, which equals
\be\label{eq:VolumeNoK}
\frac{S_{d-2}}{d(d-1)}T^d\left (1-\frac{d}{2 (d+1)}\Gamma^{0}_{\;ij} (p_0)A^{i}_{\;\ibar}A^{j}_{\;\jbar}\delta^{\ibar\jbar}T\right)
+O (T^{d+2}) \;,
\ee
where the $\delta^{\ibar\jbar}$ comes from the fact that cross terms ($\ibar\neq \jbar$) vanish under the angular integration. The defining relations for $A^{i}_{\;\ibar}$ can be rearranged to give $A^{i}_{\;\ibar}A^{j}_{\;\jbar}\delta^{\ibar\jbar}=h^{ij} (p_0)$, and in GNCs the extrinsic curvature on the surface is given by
\be\label{eq:K}
K
=g^{\mu\nu }\nabla_{\mu}n_{\nu}
=-\Gamma^{0}_{\;ij}h^{ij}=-\frac{1}{2}\frac{\dot{h}}{h} \;.
\ee
Substituting this into~\eqref{eq:VolumeNoK} we obtain 
\begin{equation} 
\frac{S_{d-2}}{d(d-1)}T^d\left (1+\frac{d}{2 (d+1)}K (0,\mathbf{x}_p)T\right)
+O (T^{d+2}) \label{eq:TopVolumeWithK} \;,
\end{equation}
and we see the first contribution is the volume of the flat cone with flat base 
as expected, and the first correction is of order $T^{d+1}$.

The corrections \textit{(i)} and \textit{(ii)} come from the non-flatness of the metric. The determinant
$\sqrt{-g}$ can be expanded in RNCs and the deviation of $T_p$ from straight lines 
considered. The curvature contribution to the volume of a small, approximately flat causal interval --
or Alexandrov neighbourhood -- of these effects has been calculated \cite{Myrheim:1978, Gibbons:2007nm, Khetrapal_Sumati:Causal_Diamond_Volume} and the same arguments show that 
the corrections \textit{(i)} and \textit{(ii)} in our case are of the same order, $O(T^{d+2})$, 
which means they are suppressed with respect to the correction derived above. This is to be expected
on dimensional grounds as extrinsic curvature has dimensions of inverse length whereas
 Riemann curvature has dimensions of inverse length squared. We will 
see that $(T^{d+2})$ corrections do not contribute to the boundary term in the limit. 

So we have
\begin{equation} 
V_\blacktriangle (T,\mathbf x)
=\frac{S_{d-2}}{d(d-1)}T^d\left (1+\frac{d}{2 (d+1)}K (0,\mathbf{x})T\right)
+O (T^{d+2}) \label{eq:TopVolumeWithK} \;.
\end{equation}
The same calculation  for $p \in U_\Sigma(\varepsilon)\cap M^-$ gives
\begin{equation}
V_\blacktriangledown (-T,\mathbf x)
=\frac{S_{d-2}}{d(d-1)}T^d\left (1-\frac{d}{2 (d+1)}K (0,\mathbf{x})T\right)
+O (T^{d+2}) \;, \label{eq:BottomVolumeWithK}
\end{equation}
where $T>0$. 

\subsection{The Mean of $\textbf{S}^{ (d)}_{CBT}$}

Focussing on $\left\langle \BP{k}\right\rangle$, we use \eqref{eq:TopVolumeWithK} in
\eqref{eq:nmax_and_eq:nmin} to find
\begin{gather}\label{eq:n_min_different_rho_terms}
\begin{aligned}
\left\langle \BP{k}\right\rangle & = \frac{\rho^{k+1}A^k}{k!}\int_{\Sigma}d^{d-1}x\: h^{\frac{1}{2}}\int_{0}^{\varepsilon}dt\:
\left[ \left ( t^{dk} + \left (kB-K \right)t^{dk+1}\right) + O\left (t^{dk+2}\right) \right] e^{-\rho A\: t^d}
 \\
 & - \frac{\rho^{k+2}A^{k+1}}{k!}\int_{\Sigma}d^{d-1}x\: h^{\frac{1}{2}}\int_{0}^{\varepsilon}dt\:
\left[ Bt^{dk+d+1} + O\left (t^{dk+d+2}\right) \right] e^{-\rho A\: t^d} + \dots \;,
\end{aligned}
\end{gather}
where $K:=K(0,\mathbf{x})$ and we have defined
\begin{gather}\label{A_and_B_defn}
\begin{aligned}
A & := \frac{S_{d-2}}{d(d-1)} \;, \\
B & := \frac{d}{2 (d+1)}K \;.
\end{aligned}
\end{gather}
To find out how this behaves in the limit of $\rho \rightarrow\infty$, it suffices to find the 
behaviour of the following integral:
\be\label{eq:general_t_n_integral}
\rho^{p}\int_{0}^{\varepsilon}dt\
t^{q}e^{-\rho At^{d}} \;,
\ee
where $p,q \in \mathbb{R}$. We make the substitution $z=\rho At^{d}$ to put~\eqref{eq:general_t_n_integral} into the form of an incomplete gamma function.
\be\label{eq:incomplete_gamma_function}
\frac{A^{-\left (\frac{q+1}{d} \right)}}{d}\rho^{p-\left (\frac{q+1}{d} \right)}\int_{0}^{\rho A \varepsilon^d}dz\
z^{\left (\frac{q+1}{d} \right)-1}e^{-z} \;.
\ee
As $\rho \rightarrow \infty$ we have 
\be\label{eq:gamma_function}
\int_{0}^{\rho A \varepsilon^d}dz\
z^{\left (\frac{q+1}{d} \right)-1}e^{-z}=
\Gamma\left ( \frac{q+1}{d} \right) + \dots \;,
\ee
where, as before, $+ \dots$ denotes terms that tend to zero exponentially fast.

Then the limiting behaviour of 
$\left\langle \BP{k}\right\rangle$ is 
\begin{gather}\label{eq:nmax_nmin_final}
\begin{aligned}
\left\langle \BP{k}\right\rangle =& \rho^{1-\frac{1}{d}} \left (b_d\right)^{-1} \frac{\Gamma\left (\frac{1}{d}+k\right)}{k!}
\int_{\Sigma}d^{d-1}x\: \sqrt{h} 
 \\
 &  -\rho^{1-\frac{2}{d}} \left (a_d\right)^{-1} \frac{\Gamma\left (\frac{2}{d}+k\right)}{k!}
\int_{\Sigma}d^{d-1}x\: \sqrt{h}K + O\left (\rho^{1-\frac{3}{d}} \right) \;, 
\\
\left\langle \BF{k}\right\rangle = & \rho^{1-\frac{1}{d}} \left (b_d\right)^{-1} \frac{\Gamma\left (\frac{1}{d}+k\right)}{k!}
\int_{\Sigma}d^{d-1}x\: \sqrt{h} 
 \\
 &   +\rho^{1-\frac{2}{d}} \left (a_d\right)^{-1} \frac{\Gamma\left (\frac{2}{d}+k\right)}{k!}
\int_{\Sigma}d^{d-1}x\: \sqrt{h}K + O\left (\rho^{1-\frac{3}{d}} \right) \;,
\\
\end{aligned}
\end{gather}
and we have given the behaviour of $\left\langle \BF{k}\right\rangle$ as well for completeness.

The results for the boundary terms and the surface volume follow almost immediately:
\begin{gather}\label{eq:boundary_final_proof}
\begin{aligned}
\lim_{\rho\rightarrow\infty} & \left\langle \textbf{S}^{ (d)}_{CBT}\right\rangle = \lim_{\rho\rightarrow\infty} \frac{\rho^{\frac{2}{d}-1}}{l_p^{d-2}} a_{d}\left (\sum_m p_m \left\langle\textbf{F}_m\right\rangle  + \sum_n q_n \left\langle\textbf{P}_n\right\rangle\right)
\\
= & \lim_{\rho\rightarrow\infty}\Bigg[\frac{\rho^{\frac{1}{d}}}{l_p^{d-1}}\frac{a_d}{b_d}\left(\sum_m p_m \frac{\Gamma\left (\frac{1}{d}+m \right)}{m!}  + \sum_n q_n\frac{\Gamma\left (\frac{1}{d}+n \right)}{n!} \right) \int_{\Sigma}d^{d-1}x\: \sqrt{h}
\\
+ &\frac{1}{l_p^{d-2}}\left(\sum_m\frac{\Gamma\left (\frac{2}{d}+m \right)}{m!}  - \sum_n q_n\frac{\Gamma\left (\frac{2}{d}+n \right)}{n!} \right) \int_{\Sigma}d^{d-1}x\: \sqrt{h}K + O(\rho^{-\frac{1}{d}})\Bigg]
\\
= & \frac{1}{l_p^{d-2}}\int_{\Sigma}d^{d-1}x\: \sqrt{h}K \;,
\end{aligned}
\end{gather}
using the conditions \eqref{coefficient_relation1} and~\eqref{coefficient_relation2} for $\vec{p}$ and $\vec{q}$. Also
\begin{gather}\label{eq:area_final_proof}
\begin{aligned}
\lim_{\rho\rightarrow\infty} & \left\langle \textbf{A}^{ (d)}\right\rangle = \lim_{\rho\rightarrow\infty} \frac{\rho^{\frac{1}{d}-1}}{l_p^{d-1}} b_{d}\left (\sum_m p_m \left\langle\textbf{F}_m\right\rangle  + \sum_n q_n \left\langle\textbf{P}_n\right\rangle\right)\\
= & \lim_{\rho\rightarrow\infty}\Bigg[ \frac{1}{l_p^{d-1}} \left(\sum_m p_m \frac{\Gamma\left (\frac{1}{d}+m \right)}{m!}  + \sum_n q_n\frac{\Gamma\left (\frac{1}{d}+n \right)}{n!} \right) \int_{\Sigma}d^{d-1}x\: \sqrt{h}+O(\rho^{-\frac{1}{d}})\Bigg]
\\
= & \frac{1}{l_p^{d-1}}\int_{\Sigma}d^{d-1}x\: \sqrt{h} \;,
\end{aligned}
\end{gather}
using~\eqref{area_coefficient_relation}. 

The form of~\eqref{eq:nmax_nmin_final}  suggests that $\left\langle \BF{k}\right\rangle$ and $\left\langle \BP{k}\right\rangle$ can be written as a Laurent series in the discreteness length, $l$, starting at $l^{1-d}$. If the 
coefficients of the higher order terms in $l$ are proportional to higher order normal derivatives of the surface volume then one might be able to find causal set analogues for second and higher normal derivatives of the surface volume.

It is interesting to note that from the limiting behaviour of $\left\langle \BP{k}\right\rangle$ or $\left\langle \BF{k}\right\rangle$ we can find the dimension, $d$, by taking the ratio of either $\left\langle \BP{0}\right\rangle$ and $\left\langle \BP{1}\right\rangle$, or $\left\langle \BF{0}\right\rangle$ and $\left\langle \BF{1}\right\rangle$. The limiting behaviour of the latter ratio, expressed in terms of the discreteness length $l$, is found to be
\begin{equation}\label{eq:dimension_estimator}
\frac{\left\langle \BF{0}\right\rangle}{\left\langle \BF{1}\right\rangle}=d-\frac{b_d \Gamma\left(\frac{2}{d} \right)}{a_d \Gamma\left(\frac{1}{d}+1 \right)}\frac{\int_{\Sigma}d^{d-1}x\: \sqrt{h}K}{\int_{\Sigma}d^{d-1}x\: \sqrt{h}}\;l+O(l^2) \;.
\end{equation}
In the limit of $l\rightarrow 0$ one gets the dimension exactly. The fraction involving the two integrals is simply the average value of the extrinsic curvature across $\Sigma$. The case for the ratio of $\left\langle \BP{0}\right\rangle$ and $\left\langle \BP{1}\right\rangle$ is the same as~\eqref{eq:dimension_estimator} but with a positive sign after $d$.

\subsection{Finite $\rho$ and Fluctuations}\label{flucts}

To decide under what circumstances the causal set boundary term, evaluated on a single causal 
set sprinkled into $M$, is close to the continuum GHY boundary term of $\Sigma$,
 it is necessary to know both the size of the 
fluctuations about the mean and when that mean is close to its limiting value. 

To take the second point first, the mean is close to its limiting value when the next order term in the 
expansions performed in the previous section can be ignored. Firstly, $\rho$ must be large enough that 
an $\varepsilon >0$ exists such that the expansions in GNCs are valid in a neighbourhood $U_\Sigma(\varepsilon)$,
and such that $\rho V_{min} \gg 1$ so $e^{-\rho V_{min}} \ll 1$, and the integral over the 
region outside $U_\Sigma(\varepsilon)$ is negligible. $V_{min} \sim \varepsilon^d$, and so $\varepsilon \gg l$. 
The expansions in equations (\ref{eq:nmax_and_eq:nmin}), (\ref{eq:TopVolumeWithK}) 
and (\ref{eq:BottomVolumeWithK}) are valid if at each point of $\Sigma$ there exist RNC compatible 
with the GNC such that 
$\mathcal{K} \varepsilon \ll 1$ and $\mathcal{R} \varepsilon^2 \ll \mathcal{K} \varepsilon$, where $\mathcal{K}$ and 
$\mathcal{R}$ stand for any component of the extrinsic curvature of $\Sigma$  and spacetime 
curvature of $M$, respectively,  evaluated on $\Sigma$. The resulting conditions are 
$ \mathcal{R} l^2 \ll \mathcal{K} l \ll 1$ which is just what one would expect. If these conditions 
did not hold then a discrete manifold with discreteness on scale $l$ could not be expected to 
encode the geometry of  $\Sigma$ and $M$ around $\Sigma$. 

We now turn to the fluctuations or standard deviation, $\sigma[\textbf{S}^{ (d)}_{CBT}]=\text{Var}[\textbf{S}^{ (d)}_{CBT}]^\frac12$, of the causal set boundary term around the mean. A heuristic argument gives an estimate of the dependence of fluctuations on $\rho=l^{-d}$. In any spacetime region of fixed volume $V$ the number of causal set elements in a sprinkling is a random variable, $\textbf{N}$, with mean $\left\langle\textbf{N}\right\rangle=\rho V$  and s.d. $\sqrt{\left\langle\textbf{N}\right\rangle}$. Consider the simplest boundary term $\textbf{S}^{ (d)}_{0}$. The volume of a region corresponding to a thickening of the hypersurface $\Sigma$ by one unit of the discreteness scale $l$ (e.g. by Lie dragging the surface along its normal by an amount $l$) is approximately $\mathrm{vol}(\Sigma) l=\mathrm{vol}(\Sigma)\rho^{-\frac{1}{d}}$. Since $\BF{0}$ and $\BP{0}$ are random variables that count nearest neighbours of $\Sigma$  we may therefore expect their mean values to scale like $\rho\mathrm{vol}(\Sigma)\l=\mathrm{vol}(\Sigma)\rho^\frac{d-1}{d}\propto\left\langle\textbf{N}\right\rangle^\frac{d-1}{d}$, and indeed this agrees with the leading order behaviour of~\eqref{eq:nmax_nmin_final}. This suggests that $\mathbf P_0$ and $\mathbf F_0$ will
be subject to fluctuations of order $\left\langle\textbf{N}\right\rangle^\frac{d-1}{2d} = (\rho V)^\frac{d-1}{2d}$
in the limit of large $\rho$. Moreover $\BF{0}$ and $\BP{0}$ are independent and so $\sigma[\textbf{S}^{ (d)}_{CBT}]$ should behave like $\rho^\frac{2-d}{d}\rho^\frac{d-1}{2d}=\rho^\frac{3-d}{2d}$. Hence for $d=2$ these fluctuations should grow like $\rho^{\frac{1}{4}}$ as $\rho\rightarrow\infty$, for $d=3$ they should be constant, and for $d>3$ they should be damped.

We tested this with simulations in the simplest case of 
flat spacetime and flat surface $\Sigma$. We took a sample of 100, density $\rho=l^{-d}$, sprinklings
of a $d$-cube $[0,1]^d$ in $d$-dimensional Minkowski space
with  hypersurface $\Sigma: t=1/2$, and evaluated the sample mean and (corrected) sample standard deviation
of $\mathbf S^{ (d)}_0$. 
The expectation value of $\mathbf S^{ (d)}_0$ is exactly zero due to the symmetry of the situation.
\begin{figure}[t!]
  \centering
    {\includegraphics[scale=0.6]{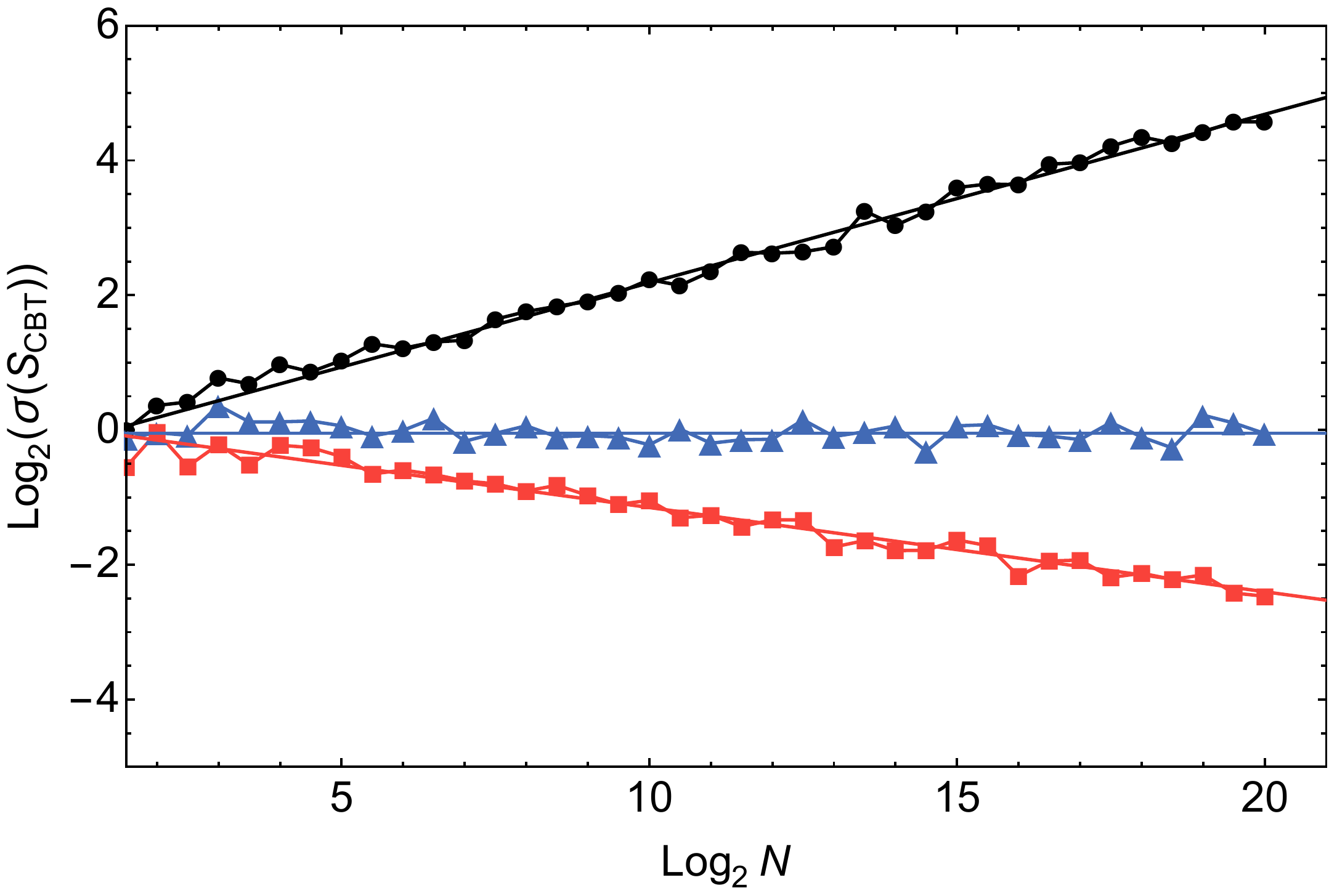}}
     \caption{A  plot  of the standard deviation in samples of 100 of $\mathbf S^{ (d)}_{0}$ for a flat ($K=0$) surface
     bisecting a $d=2,3$ and $4$-dimensional unit cube in Minkowski space for different values of $N=\rho$. 
     Black dots, blue triangles and red squares correspond to the simulation results in $d=2,3$ and $4$ dimensions, respectively. The corresponding black, blue and red lines have gradients $\frac14$, $0$ and $-\frac18$ and best-fit intercepts of order $1$.}
     \label{fig:fluctuations}
\end{figure}

Figure~\ref{fig:fluctuations} shows  the results for $d=2,3,4$ spacetime dimensions, with $\left\langle\textbf{N}\right\rangle=\rho$ ranging up to $2^{20}$. Each data point represents the 
sample standard deviation for a sample of 100. The solid lines have been obtained by fitting an arbitrary constant multiplier in the scaling law predicted by the argument above, $\gamma (d)\times \left\langle\textbf{N}\right\rangle^\frac{3-d}{2d}$, to the data. The best fit values are all of order $1$: $\gamma (2)=0.80$, $\gamma (3)=0.97$, and $\gamma (4)=1.07$.
The data are evidence for the scaling predicted by the heuristic argument. The sample means (not shown) for different $\rho$ are consistent with zero within the standard error.
Simulations for the boundary term $\mathbf S^{ (d)}_+$ (which is proportional to $d \mathbf F_1 - \mathbf F_0$) show the same dimension dependent 
scaling behaviour for the standard deviation, though in this case the heuristic argument is complicated by the fact that the random variables of which the boundary term is a sum are not independent.%
\footnote
{While the heuristic argument predicts a scaling of the mean and standard deviations consistent with the data, a closer look at the samples we generated for $\mathbf F_k$ and $\mathbf P_k$ for $k=0,1$ suggests that their distributions deviate from a Poisson distribution: they are ``underdispersed'', i.e. their s.d. grows like the square root of the mean but is related to it by a constant of proportionality less than 1. We have begun to investigate this further and hope to return to a more careful study of the distributions of these random variables in a future note.}

\section{The Causal Set Action for a Flat Alexandrov Interval}\label{intervals}
\newcommand{\vol}{\mathrm{vol}}
\newcommand{\tS}{\textbf{S}}
\newcommand{\tN}{\textbf{N}}
\newcommand{\la}{\langle} 
\newcommand{\ra}{\rangle} 
\newcommand{\jv}{\mathcal {J}}
\newcommand{\sA}{\mathcal {A}}

Now that we have a family of analogue GHY boundary terms for causal sets in hand we can consider if such terms need to be 
included in any putative action for causal sets.  In particular  we can ask whether boundary terms need to be
added to the recently proposed Benincasa-Dowker-Glaser (BDG) causal set actions \cite{Benincasa_Dowker:The_Scalar_Curvature_of_a_Causal_Set, Dowker_Glaser:dAlembertians_for_Causal_Sets, Glaser:2013xha}. 
Before that question can be answered, it is necessary to determine whether the BDG actions already contain any boundary contributions. 

The BDG action  $S_{BDG}^{ (d)}\left[\mathcal C\right]$  of a finite causal set $\mathcal{C}$ is
\begin{equation} 
\frac{1}{\hbar}{S_{BDG}^{ (d)}}\left[\mathcal C\right] = -\alpha_d (l/l_p)^{d-2} \biggl ( N[\mathcal C]+ \frac{\beta_d}{ \alpha_d} \sum_{i=1}^{n_d-1} C^{ (d)}_{i} N _i[\mathcal C] \biggr) \;, \label{bd} \end{equation} 
where $N_i[\mathcal C]$ is the number of $(i+1)$-element inclusive order intervals in $\mathcal{C}$,
$N[\mathcal C]$ is the cardinality of the causal set, and $l/l_p$ is the ratio of a
fundamental length to the Planck length\footnote{We reintroduce $\hbar$ in this section.}.   The constants are 
\begin{equation} \begin{aligned}
 \alpha_d =   
\begin{cases}
\displaystyle
-\frac1{\Gamma\left (1+\frac2d\right)}c_d ^{2/d} \;   &d \, \, \mathrm{odd} \\
\displaystyle
- \frac{2}{\Gamma\left (1+\frac2d\right)}c_d ^{2/d}  \;   &d \, \,  \mathrm{even} \;,\\
\end{cases}
%\quad\text{and}\quad 
\end{aligned}
\end{equation}
\begin{equation}
\begin{aligned}
\beta_d =
\begin{cases}\displaystyle
\frac{d+1}{2^{d-1}\Gamma\left (1+\frac2d\right)} c_d^{2/d}   \;   &d\mathrm{ \, \, odd}\\ 
\displaystyle
\frac{\Gamma\left (\frac{d}{2}+2\right)\Gamma\left (\frac{d}{2}+1\right)}{\Gamma\left (\frac{2}{d}\right)\Gamma\left (d\right)}c_d^{2/d}  &  d\mathrm{ \, \,  even} \;,\\ 
\end{cases} 
\end{aligned}
\end{equation}
and
\begin{equation} 
n_d = 
\begin{cases} 
\frac{d}{2} + \frac{3}{2}  \quad & d\mathrm{ \, \, odd}\\ 
\frac{d}{2} + 2  \quad &d\mathrm{ \, \,  even} \;,\\ 
\end{cases} 
\end{equation} 
where $c_d= 2^{1 -\frac{d}{2}}S_{d-2}/(d(d-1))$ (recall that $S_d$ is the volume of the unit $d$-sphere).
The coefficients $C_i^{ (d)}$ of the terms $N_i[\mathcal C]$ in the sum are
\begin{equation}
\label{cid}
 C_i^{ (d)}= 
\begin{cases} 
\displaystyle\sum_{k=0}^{i-1} (-1)^k\binom{i-1}{k} \frac{\Gamma\left(\frac{d}2(k+1)+\frac32\right)}{\Gamma\left (\frac{d}{2}+ \frac{3}{2}\right) \Gamma\left (\frac{d}2k + 1\right)}  \quad  &d\mathrm{ \, \, odd}\\ 
\displaystyle\sum_{k=0}^{i-1} (-1)^k\binom{i-1}{k} \frac{\Gamma\left (\frac{d}2 (k+1)+2\right)}{\Gamma\left (\frac{d}2+2\right) \Gamma\left (\frac{d}2k+1\right)}  &d \mathrm{ \, \,  even} \;.\\ 
\end{cases} 
\end{equation} 
We note here that these coefficients can be expressed more compactly as generalised hypergeometric functions of type $\{ q+1,q\}$:
\begin{equation}
C_i^{(d)}={}_{q+1}F_{q} \left(\{a_1, \ldots, a_q, i-1\}, \{b_1, \ldots, b_q\}|1\right) \;,
\label{chyp} 
\end{equation} 
with $q=\frac{d+1}{2}$, $a_i=\frac{d+2i}{d}$ and $b_i=\frac{2i}{d}$ for $d$ odd, and $q=\frac{d}{2}$, $a_i=\frac{d+2i+2}{d}$ and $b_i=\frac{2i}{d}$ for $d$ even.  

As in Section~\ref{claims}, given a causal Lorentzian spacetime $(M, g)$, the sprinkling process at 
density $\rho = l^{-d}$ turns this function of  causal sets into a random variable 
 $\textbf{S}^{ (d)}_{BDG}[M, \rho]$, the ``random discrete action'' of $(M, g)$.
 A requirement for the causal set action to be physically interesting is that its mean should
 tend to the continuum action of $(M, g)$  as $\rho \rightarrow \infty$. The question at hand
 is whether  in this limit it includes boundary contributions in addition to the Einstein-Hilbert term. 

\begin{figure}[t!]
 \centering
   {\includegraphics[scale=0.6]{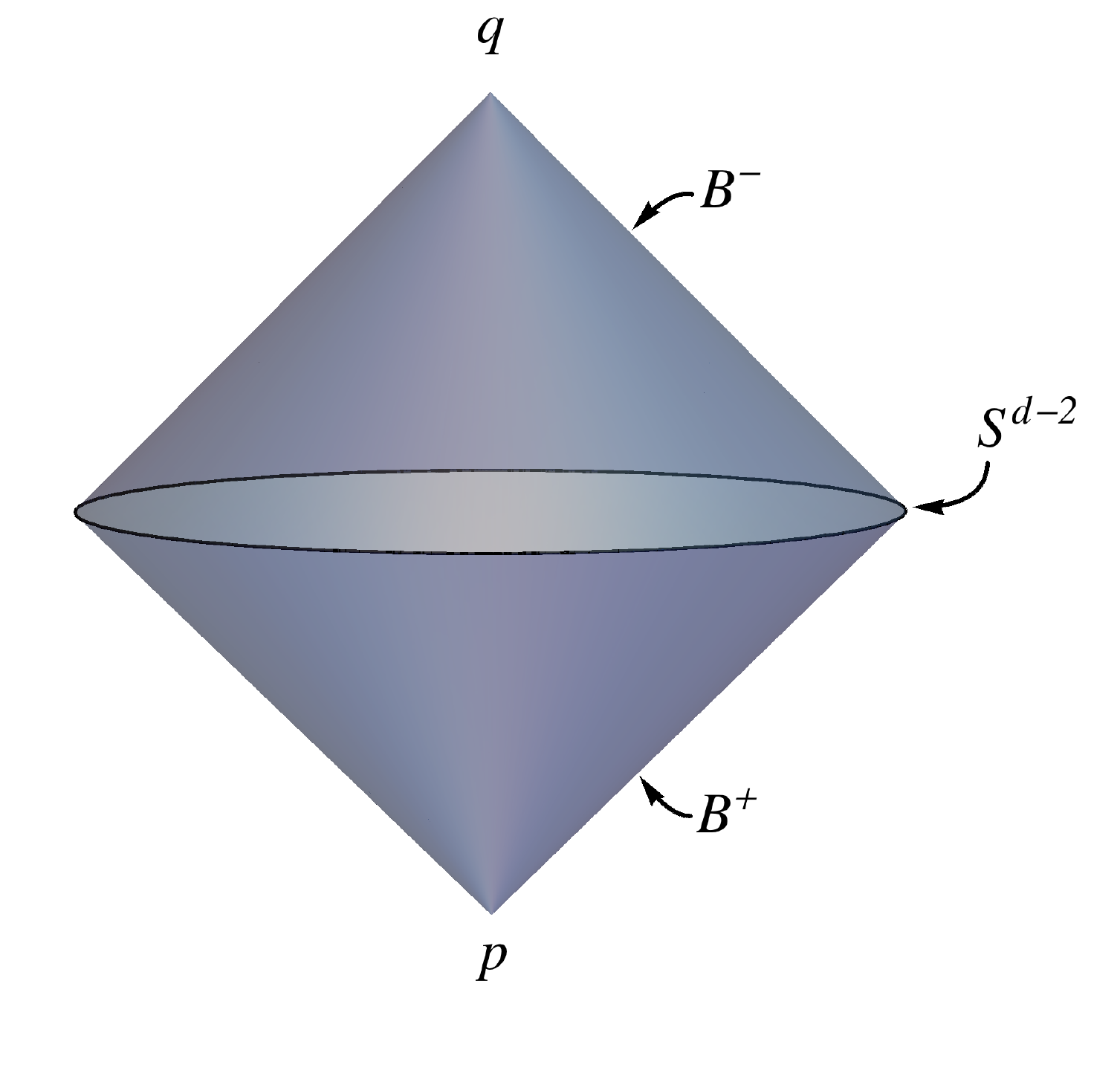}}
    \caption{The Alexandrov interval $I(p,q)$. The boundary consists of the null sections $B^\pm$ and the spatial sphere $S^{d-2}$ at their joint.} 
    \label{fig:Alex}
\end{figure}

We will explore this question by calculating the mean of the $d$-dimensional BDG action for causal sets sprinkled into causal intervals or ``Alexandrov intervals'' in  $d$-dimensional flat spacetime. Since the Einstein-Hilbert 
contribution is expected to be zero, this will teach us something about what boundary contributions, if any, are included in the BDG action. The boundary of an Alexandrov interval consists of a past and a future null cone which intersect at
a codimension-2 \textit{joint} of topology $S^{d-2}$ (see Figure~\ref{fig:Alex}). 
 While the GHY term is defined in the
continuum for both spacelike and timelike boundaries, and contributions from their codimension-2
intersections or joints have also been worked out \cite{Hayward:1993my}, null boundaries and their
intersections, on the other hand, are little discussed and there is no consensus on whether the
variational principle for GR can be made well-defined on a region with null boundaries (see,
however, \cite{neiman,paddy}).

As was shown in \cite{Benincasa:2010as}, when $d=2$ 
the continuum limit of the expectation value of the discrete random action of an Alexandrov interval 
of arbitrary size is equal to 2.
While this might suggest topological invariance, we will now show that it is a part of a more general result for $d>2$ and has a geometrical origin. Namely, it corresponds to the volume of the joint of the Alexandrov interval, which in flat spacetime is a $(d-2)$-sphere and independent of the interval size {\it only} in $d=2$. 

Consider an Alexandrov interval, $I(p,q)$, of proper height $\tau$ between two points $p$ and $q$ in
$d$-dimensional Minkowski spacetime.  Its boundary consists of the two null cones from $p$ and $q$
which intersect at the joint, $\mathcal{J}^{(d-2)}:=\partial J^+ (p) \cap \partial J^- (q)$, a
codimension-2 sphere of radius $\tau/2$.  The joint has volume ${\mathrm{vol}}(\mathcal{J}^{(d-2)})
= (\tau/2)^{d-2} S_{d-2}$. The interval itself has volume ${\mathrm{vol}} (I (p,q)) = 2 (S_{d-2}/(d
(d-1))) (\tau/2)^{d}$. For the sprinkling process at density $\rho = l^{-d}$, the mean,
$N:=\left<\mathbf N\right>$, of the number of causal set elements sprinkled into $I (p,q)$ is $ N =
\rho\, {\mathrm{vol}} (I (p,q)) $. In what follows we take the continuum limit $\rho \rightarrow
\infty$ while keeping $\tau$ fixed.  The mean of the random discrete action of this
flat region should give, in the limit of large $\rho$, contributions from the boundary only.

In \cite{Glaser_Sumati:Locality_in_Causal_Set} a closed form expression was obtained for the mean
value of the number  of $(i+1)$-element inclusive intervals  contained in an Alexandrov interval in $d$-dimensional flat spacetime:
\begin{equation} \left \langle \tN_i^{(d)}\right \rangle\!=\! \frac{\Gamma\left (d\right)^2 N^{i+2}}{\Gamma\left
      (i\right)} \sum_{k=0}^\infty \frac{ (-N)^k\,\Gamma\left(k+i+1\right)\Gamma\left (\frac{d(k+i)}{2}+1\right)  \Gamma\left
      (\frac{d(k+i+1)}{2}+1\right)}{ \Gamma\left(k+i+3\right) \Gamma\left (k+1\right)\Gamma\left (\frac{d
      (k+i)}{2} +d\right) \Gamma\left (\frac{d(k+i+1)}{2}+d\right)} \;, \label{abundances}
\end{equation} 
where $i \geq 1$. Importantly, this power series can be expressed more compactly in terms of  a generalised hypergeometric function of type $\{d,d\}$ as shown in 
\cite{Glaser_Sumati:Locality_in_Causal_Set}, and is therefore convergent for all $N$. All the  power series in $N$ that appear subsequently in this section are therefore also convergent.
 We now use this to evaluate $\langle \tS_{BDG}^{ (d)}\rangle $ in an
Alexandrov interval in flat spacetimes of different dimensions.

We begin with the simplest case of $d=2$, where 
\begin{equation}
 \frac{1}{\hbar}\left \langle \tS_{BDG}^{ (2)} \right\rangle = 2 \biggl ( N -2 \left \la \tN_1^{(2)}
 \right\ra +4 \left \la
 \tN_2^{(2)}\right\ra -2 \left \la \tN_3^{(2)} \right \ra \biggr) \;.
\end{equation} 
Using (\ref{abundances}) gives  a power series expansion in $N$ with coefficients  
$\frac{(-1)^{i-1}}{i!} $,  $i=1, \ldots \infty $, so that 
\begin{equation} 
 \frac{1}{\hbar}\left \langle \tS_{BDG}^{ (2)} \right\rangle = 2\left( 1 - e^{-N}\right) \;,
\end{equation} 
which agrees with the result in \cite{Benincasa:2010as}. In anticipation of the  results for higher $d$  we note that
the volume of the zero sphere at the joint,   ${\mathrm{vol}}(\mathcal{J}^{(0)})
= S_{0}=2$, so that 
\begin{equation}
\lim_{N \rightarrow {\infty}}  
 \frac{1}{\hbar}\left\langle \tS_{BDG}^{ (2)} \right\rangle =
{\mathrm{vol}}(\mathcal{J}^{(0)}) \;.
\end{equation}  
This is in agreement with the result obtained for a 2-dimensional flat  causal interval \cite{Benincasa:2010as}. 

Next, substituting (\ref{abundances}) into the $d=3$  averaged BDG action,  
\begin{equation} 
 \frac{1}{\hbar}\left \langle \tS_{BDG}^{ (3)}\right \rangle = -\alpha_3
 \biggl(\frac{l}{l_p}\biggr)\biggl ( N - \left\la \tN_1^{(3)} \right \ra + \frac{27}{8} \left \la
 \tN_2^{(3)}\right \ra -\frac{9}{4} \left\la \tN_3^{(3)} \right \ra \biggr) \;,
\end{equation} 
gives a power series expansion  in $N$ with coefficients 
\begin{equation} 
-\alpha_3  \biggl(\frac{l}{l_p}\biggr)\times \frac{(-1)^{i+1} }{i!}\frac{8}{(3i+1) (3i-1)} \;,
\end{equation}  
where $i=1, \ldots \infty $.  Rearranging indices we find a  closed form for the action:  
\begin{equation} 
 \frac{1}{\hbar}\left \langle \tS_{BDG}^{ (3)} \right \rangle = - 8 \alpha_3 \biggl(\frac{l}{l_p}\biggr) \times
 \biggl( -1 +  {}_2F_{2}\left.\left(\left\{\text{\footnotesize $\frac{1}{3}$},
 \text{\footnotesize $-\frac{1}{3}$}\right\},\left\{\text{\footnotesize $\frac{4}{3}$},\text{\footnotesize $\frac{2}{3}$}\right\} \right|-N \right) \biggr) \;,
\end{equation}
where ${}_2F_{2}$ is a generalised hypergeometric function of type $\{2,2\}$. This can be
re-expressed more simply as 
\begin{equation} 
 \frac{1}{\hbar}\left \langle \tS_{BDG}^{ (3)} \right \rangle = - 8 \alpha_3 \biggl(\frac{l}{l_p}\biggr)
 \biggl(  - 1 + \frac{1}{6N^{\frac{1}{3}}} \gamma\left(\frac{1}{3}, N\right) - 
   \frac{N^{\frac{1}{3}}}{6} \gamma\left(-\frac{1}{3},N\right) \biggr) \;,
\label{simpthree}
\end{equation} 
where  $\gamma(s,x)\equiv \int_0^x t^{s-1}e^{-t}  dt$  is a lower  incomplete Gamma function. 
The large $N$ behaviour is thus dominated by the last term in the above expression.  Using
$\gamma(s,x) = \Gamma(s)-\Gamma(s,x)$, where the upper incomplete Gamma function $\Gamma(s,x)
\sim x^{s-1}e^{-x}$ in the asymptotic limit, the dominant term in (\ref{simpthree}) simplifies to
$-4\alpha_3 l  N^{1/3} \Gamma(2/3)/l_p =\vol(\jv^{(1)})/ l_p$. Hence
\begin{equation}
\lim_{N \rightarrow {\infty}}  
 \frac{1}{\hbar}\langle \tS_{BDG}^{ (3)} \rangle = \frac{1}{l_p}  \vol(\jv^{(1)}) \;. 
\end{equation}  

For $d=4$
\begin{equation} 
 \frac{1}{\hbar}\left\langle \tS_{BDG}^{ (4)} \right\rangle = -\alpha_4 \biggl(\frac{l}{l_p}\biggr)^2\biggl ( N -
 \left\la \tN_1^{(4)} \right\ra + 9  \left\la
 \tN_2^{(4)}\right\ra -16 \left\la \tN_3^{(4)} \right\ra + 8 \left\la \tN_4^{(4)} \right\ra\biggr) \;.
\end{equation} 
Excluding the first term, this is  a power series in $N$ with coefficients
\begin{equation}
  -\alpha_4 \biggl(\frac{l}{l_p}\biggr)^2 \times \frac{(3!)^2}{3} \frac{ (-1)^{i+1} (i-1)(2i-3)!}{i!(2i+1)!} \;,
\end{equation}    
where now $i=2,\ldots, \infty$. Using this,   \textit{Mathematica} yields the closed form expression
\begin{gather} 
\begin{aligned} 
 \frac{1}{\hbar}\left\langle \tS_{BDG}^{ (4)} \right\rangle = & -\alpha_4 \left(\frac{l}{l_p}\right)^2 \biggl(
\frac{3 (2N-1)}{2\sqrt{N}} \sqrt{\pi} \mathrm{Erf} \left(\sqrt{N}\right)
\\
& - 3 \left( \gamma - e^{-N}+
 \Gamma(0,N)+\ln(N)\right)\biggr) \;,
\end{aligned}
\end{gather}
where $\gamma$ is the Euler-Mascheroni constant, and $\mathrm{Erf} $ is the error function. Since
$\mathrm{Erf}(\sqrt{N}) $ goes to $ 1$ in the asymptotic limit, the dominant contribution to the
above expression comes from the second term, $ -3 \alpha_4 l^2 \sqrt{\pi N}/l_p^2$, which  simplifies
to $2\sqrt{6 \pi N} l^2/l_p^2 = \vol(\jv^{2})/l_p^2 $.  Thus, again
\begin{equation}
\lim_{N \rightarrow {\infty}}  
 \frac{1}{\hbar}\left\langle \tS_{BDG}^{ (4)} \right\rangle =\frac{1}{l_p^2} 
{\mathrm{vol}}(\mathcal{J}^{(2)}) \;. 
\end{equation}  

We now turn to the case of general $d$. We begin by  writing the (averaged) sum in  (\ref{bd}) as a power series in $N$:
\be
\sum_{i=1}^{n_d}C_i^{ (d)} \left\langle \tN_i^{}\right\rangle = \sum_{j=1}^\infty A_j^{ (d)}
N^{j+1} \;.
\label{powseries}
\ee
After a rearrangement and redefinition of indices we find that  
\begin{equation}
 A_j^{ (d)} = \Gamma\left (d\right)^2\frac{ (-1)^j}{ (j+1)!}\frac{\Gamma\left (\frac{d}{2}
     (j-1)+1\right)\Gamma\left (\frac{d}{2}j+1\right)}{\Gamma\left (\frac{d}{2} (j-1)+d\right)
   \Gamma\left (\frac{d}{2}j+d\right)}\sum_{i=1}^{D+2} (-1)^i \binom{j-1}{i-1}  C_i^{ (d)} \;,
\label{Ajds}
\end{equation} 
where $d=2D$ for $d$ even and $d=2D+1$ for $d$ odd.  While (\ref{powseries}) can be directly
evaluated by \textit{Mathematica} for small values of $d=2,\ldots, 5$, it is greatly assisted by the following
simplifications to the $A_j^{(d)}$ for higher $d$.  

We begin by evaluating the sum in (\ref{Ajds}).  We first use \textit{Mathematica} to evaluate it for  $d =
2,\ldots, 20 $ which then suggests the  general form 
\begin{equation} 
\sum_{i=1}^{D+2} (-1)^i \binom{j-1}{i-1}  C_i^{ (d)} = 
\begin{cases} 
\displaystyle\frac{(-1)^D\left((2D+1)^2j^2-1\right)(3-{(2D+1)j}/2)_{D-1} }{4\Gamma(2+D)}\,   & d \, \, \mathrm{odd} \\
\displaystyle\frac{(-1)^D Dj(2+2D)(1-Dj)_{D-1}}{2\Gamma(2+D)}\,  & d \, \, \mathrm{even} \;, \\
\end{cases} 
\end{equation}  
where $(a)_k$ is the Pochhammer symbol. Inserting this into (\ref{Ajds}) we  use 
\textit{Mathematica} to evaluate it for $d= 2,\ldots, 20 $.  After some manipulations this suggests the
general expression 
\begin{equation}
A_j^{(d)}=   \frac{\Gamma\left (d\right)^2 (-1)^{j+1}}{\Gamma\left (\frac{d}{2} (j + 1)\right) \Gamma\left (\frac{d}{2} (2 + j)\right) \Gamma\left (2 + j\right) } \gamma_j^{ (d)} \;,
\label{simplercoefft} 
\end{equation} 
where 
\begin{equation} \gamma_j^{ (d)}= 
\begin{cases} 
\displaystyle\frac{\sqrt{\pi}}{2^{1+dj} }\frac{\Gamma\left (2+dj\right)}{\Gamma\left (\frac{d-1}{2}\right)}    \quad   & d \, \, \mathrm{odd} \\
\displaystyle\frac{\Gamma\left (1+\frac{d}{2}j\right)  \Gamma\left (2+\frac{d}{2}j\right) }{\Gamma\left (\frac{d}{2}\right)}  \quad  & d\mathrm{ \, \,  even} \;. \\
\end{cases} 
\end{equation} 

Taking our cue from the behaviour of $\langle \tS_{BDG}^{ (d)}\rangle$ for $d=2,3,4$ in the  $N \rightarrow \infty$ limit, we will consider the ratio
\begin{equation} 
\frac{ l_p^{d-2} \left\langle \tS_{BDG}^{ (d)}\right\rangle}{\hbar \,\,\vol (\mathcal{J}^{(d-2)})} = \frac{\varepsilon_d}{d (d-1) \Gamma\left (1+\frac2d\right) N^{\frac{d-2}{d}}} \left ( N + \frac{\beta_d}{\alpha_d} \sum_{i=1}^{n_d} C_i^{ (d)} \left\langle \tN_i \right\rangle \right) \;,
\label{ratio} 
\end{equation} 
where $\varepsilon_d=1$ for $d$ odd and $2$ for $d$ even.  Finally inserting  (\ref{simplercoefft}) into
(\ref{ratio})  \textit{Mathematica} gives  for $d=2, \ldots, 16$
\begin{equation} 
\lim_{N \rightarrow \infty} \frac{1}{\hbar}\left\langle \tS_{BDG}^{(d)} \right\rangle=  \frac{1}{ l_p^{d-2}} \vol (\mathcal{J}^{(d-2)}) \;.
\label{result}
\end{equation}
This is the main result of this section and can be interpreted as saying that, in the continuum limit, the mean of the random discrete action of a causal diamond is a pure boundary term coming only from the volume of the codimension-2 joint. One might speculate that the BDG action for the Alexandrov interval contains {\it all}
the GHY contributions which implies  that the null boundary GHY term vanishes identically.  Interestingly, this concides with the claim in \cite{neiman} that the GHY term for an Alexandrov interval is given only by the volume of the spatial joint. The result we have obtained is for flat spacetime and it would be interesting to see how the presence of curvature affects it by repeating this calculation in RNCs to the lowest order corrections.

Finally, while efforts have been made to find a closed form expression of
$\frac{1}{\hbar}\langle \tS_{BDG}^{(d)} \rangle$
 for arbitrary $d$ this has proved difficult,
even in the asymptotic limit. As we now show, the most obvious approach of using the 
asymptotic form of the $\langle \tN_i^{(d)} \rangle$s is insufficient for this purpose.
In the large $N$ limit \cite{Glaser_Sumati:Locality_in_Causal_Set}
\begin{equation}
\left\la \tN_i^{(d)}\right\ra= \frac{\Gamma\left(\frac{2}{d}+i\right)\Gamma(d)}{i!\left(\frac{d}{2}-1\right)\left(\frac{d}{2}+1\right)_{d-2}}
  N^{2-\frac{2}{d}}+ O(N^\alpha f(N)) \;,
\label{leadingone}
\end{equation}
where $\alpha=1$ for $d=3,4$ and $2-\frac{4}{d}$ for $d>4$, and $f(N)=\ln N$ for $d=4$ and $1$
otherwise. For $d=2$  
\begin{equation} 
\left\la \tN_i^{(2)}\right\ra= N \ln N +O(N) \;.
\label{leadingtwo}  
\end{equation} 
Since this dominates the leading order contribution of $N^{\frac{d-2}{d}}$ to $\frac{1}{\hbar}\langle
\tS_{BDG}^{(d)}\rangle$ for all $d$, it is clear that this contribution must vanish. Inserting
(\ref{leadingone}) and (\ref{leadingtwo}) into the BDG action  confirms that this is indeed the
case for $d=2, \ldots 16$. In fact, the next to leading order terms in (\ref{leadingone}) and
(\ref{leadingtwo}) also do not have the requisite $  N^{\frac{d-2}{d}}$ dependence, and are 
dominant in comparison. Hence their contribution too should vanish, but we do not have an explicit
expression for their coefficients to check this. Suffice to say that the asymptotic behaviour  of $\langle
\tN_i^{(d)} \rangle$  is indeed not enough to find the leading order 
dependence of the BDG action in the flat spacetime interval. 

\section{Summary}

We have found a family of causal set boundary terms that agree in the mean with the Gibbons-Hawking-York boundary term for a spacelike hypersurface. We presented a heuristic argument for how the fluctuations of these boundary terms go with $\rho$, and provided numerical evidence in $2$, $3$ and $4$ dimensions to support that reasoning. In $4$ dimensions the fluctuations decrease with $\rho$. It would be interesting to also find causal set analogues of the GHY boundary term for timelike boundaries. The situation is more complicated in that case because the identification of ``nearest neighbours" to a timelike hypersurface in terms of causal structure is less straight-forward than in the spacelike case. 

The other major result of this paper is that the average over sprinklings of the BDG action for an interval in Minkowski spacetime is proportional to the volume of the ``joint" of that interval. There is still more work to be done to determine what sort of boundary contributions are already contained in the BDG action.

Other interesting results obtained along the way were causal set expressions for the spatial volume of a spacelike hypersurface and the dimension of the manifold the causet has been sprinkled into. 

\section*{Acknowledgements}
The work was supported by STFC grant ST/J0003533/1. M.B. is supported by NSF Grant No.\ CNS-1442999. I.J. is supported by the EPSRC. S.S. is funded in part, under an agreement with Theiss Research, by a grant from the Foundational  Questions Institute (FQXI) Fund,  a donor advised fund of the Silicon Valley Community Foundation on the basis of proposal FQXi-RFP3-1346 to the Foundational Questions Institute. 

\newpage

\bibliographystyle{jhep}

\bibliography{paper}

\end{document}